# Mid-infrared continuous-filtering Vernier spectroscopy using a doubly resonant optical parametric oscillator

Amir Khodabakhsh[1], Lucile Rutkowski[1], Jérôme Morville[2], and Aleksandra Foltynowicz[1*]

[1]*Department of Physics, Umeå University, 901 87 Umeå, Sweden*
[2]*Institut Lumière Matière, CNRS UMR5306, Université Lyon 1, Université de Lyon, 69622 Villeurbanne CEDEX, France*

*email: aleksandra.foltynowicz@umu.se*

We present a continuous-filtering Vernier spectrometer operating in the 3.15-3.4 μm range, based on a femtosecond doubly resonant optical parametric oscillator, a cavity with a finesse of 340, a grating mounted on a galvo scanner and two photodiodes. The spectrometer allows acquisition of one spectrum spanning 250 nm of bandwidth in 25 ms with 8 GHz resolution, sufficient for resolving molecular lines at atmospheric pressure. An active lock ensures good frequency and intensity stability of the consecutive spectra and enables continuous signal acquisition and efficient averaging. The relative frequency scale is calibrated using a Fabry-Perot etalon or, alternatively, the galvo scanner position signal. We measure spectra of pure $CH_4$ as well as dry and laboratory air and extract $CH_4$ and $H_2O$ concentrations by multiline fitting of model spectra. The figure of merit of the spectrometer is $1.7\times10^{-9}$ cm$^{-1}$ Hz$^{-1/2}$ per spectral element and the minimum detectable concentration of $CH_4$ is 360 ppt Hz$^{-1/2}$, averaging down to 90 ppt after 16 s.

## 1. Introduction

Advances in mid-infrared (MIR) frequency comb sources have made possible broadband high resolution absorption spectroscopy in the molecular fingerprint region [1, 2] allowing simultaneous multispecies detection [3] and precision line position measurements [4]. The coherence of comb sources enables measurements over long distances [5] and efficient coupling to multipass cells [3] and cavities [6] to enhance the interaction length with the sample. The high absorption sensitivity in combination with the strong intensities of the fundamental transitions translates into low concentration detection limits for many molecular species. Still, the number of demonstrations of MIR comb spectroscopy for practical applications is limited, mostly because of the complexity of the comb sources and detection systems.

Most of the previous implementations of MIR comb spectroscopy use detection systems based either on Fourier transform spectroscopy or virtually imaged phased array (VIPA). Fourier transform spectroscopy (FTS) can be implemented either with a mechanical interferometer [3, 7-11] or using the dual-comb technique [4, 12, 13]. Both approaches enable measurements with frequency resolution and precision provided by the comb with no influence of the instrumental line shape and are suitable for broadband precision spectroscopy [4, 14]. Mechanical MIR FTS has been successfully combined with enhancement cavities [6, 15] and enabled acquisition of rotationally resolved spectra of cold complex molecules [16]. The use of the cavities with mechanical FTS requires a high-bandwidth lock and the transmitted spectral range is limited by the cavity dispersion [6]. Moreover, the measurement time is of the order of a second, fundamentally limited by the movement of the mirror, and the mechanical interferometer is a rather bulky detection system. Dual comb spectroscopy enables a faster measurement without moving parts, although reaching the same signal-to-noise ratio (SNR) as with a mechanical FTS fundamentally requires the same averaging time [17]. Moreover, to achieve a fully resolved and accurate performance, a tight phase-lock of the two comb sources is required [4]. When the comb line resolution is not needed, the requirement of the tight lock is released and a relatively stable operation of two MIR combs can be achieved e.g. by seeding one optical parametric oscillator (OPO) cavity with two pump lasers [18] or combining two OPO cavities into one [19]. Recently, MIR comb sources other than mode-locked lasers, such as quantum cascade lasers [20] and electro-optically modulated continuous wave lasers [21], have enabled compact dual comb spectrometers. Still, MIR dual comb spectroscopy with cavity enhancement has not yet been demonstrated.

An alternative detection method is to disperse the comb spectrum using a VIPA in combination with a grating and record the resulting 2D image with a camera [22]. The resolution of VIPAs is usually on the order of a few hundreds of MHz, which allows resolving the lines of high repetition rate combs [22]. For lower repetition rates, the comb lines are not resolved and the spectra require frequency calibration using the position of known absorption lines [5, 23]. VIPA-based systems allow fast measurements (on the order of ms), but the simultaneous spectral coverage is limited by the detector size, so recording of the entire comb bandwidth requires sequential measurement and changing the positions of the grating [5]. Moreover, retrieving absorption spectra from the camera images requires rather complex signal analysis [5, 15, 23] and MIR VIPAs and cameras are quite expensive. However, VIPA-based systems can rather easily be combined with enhancement cavities using a low-bandwidth dither lock, and the high sensitivity and short acquisition time have made possible time-resolved measurements of transient radicals [23] and reaction kinetics [24]. Open-air broadband trace gas sensing using a VIPA-based spectrometer has also been demonstrated [5].

Recently, a new detection method for comb spectroscopy has been proposed and demonstrated in the near-infrared wavelength range, called continuous-filtering Vernier spectroscopy (CF-VS) [25, 26], that allows the acquisition of broadband comb spectra with medium to high resolution in measurement time of the order of a few tens of ms using a compact and robust detection system. In CF-VS the comb is filtered using an external cavity whose free spectral range ($FSR_C$) is slightly detuned from the comb repetition rate ($f_{\mathrm{rep}}$). For small $FSR_C$-$f_{\mathrm{rep}}$ mismatch groups of comb lines, called Vernier orders, are transmitted through the cavity and dispersed with a grating. To record the spectrum, the intensity of a selected order is measured using a photodetector while the order is tuned across the comb spectrum by tuning of the cavity length (or $f_{\mathrm{rep}}$). The transmitted bandwidth is limited only by the high reflectivity range of the cavity mirrors rather than by their dispersion or by the size of the detector array. The principle of continuous filtering requires that Vernier orders consist of more than one comb line, which implies that resolving the comb lines is not possible and external frequency calibration is required. However, because half of the comb lines defining the Vernier order are transmitted on the positive slope of the cavity modes, and half on the negative slope, the frequency to amplitude noise conversion is drastically reduced and the technique provides high absorption sensitivity [26]. CF-VS thus enables fast and sensitive measurement of broadband comb spectra directly in the frequency domain using low bandwidth photodetectors and the signal does not require any processing as is the case for camera images in VIPA-based systems. The molecular spectra are affected by the interaction of the comb lines with the cavity modes that shift because of molecular dispersion. However, a theoretical model that accounts for this has been developed and verified, allowing quantitative measurements [26].

Note that the CF-VS technique is different than the comb-resolved Vernier spectroscopy demonstrated by Gohle *et al*. [27] where larger FSR-$f_{rep}$ mismatch is used and single comb lines are transmitted in each Vernier order. The comb-resolved VS requires higher resolving power in cavity transmission because the Vernier orders are much closer to each other in the frequency domain compared to the CF-VS regime. The technique allows precision spectroscopic measurements of transition frequencies and linewidths with the individual comb lines [28], but unless the comb source and the cavity are well stabilized, it suffers from rather poor detection sensitivity [29]. Comb-resolved VS has been demonstrated in the MIR range, but the cavity was used only as a filter and the sample was placed outside the cavity, thus not benefiting from the enhancement of the interaction length [30].

We have recently implemented the CF-VS technique in the MIR range (3.15-3.4 μm) using a spectrometer based on a doubly resonant optical parametric oscillator (DROPO) with an orientation-patterned GaAs crystal pumped by an Tm:fiber femtosecond laser, an enhancement cavity, a diffraction grating, and a photodetector [10]. In this proof-of-principle experiment we measured a spectrum of laboratory air at atmospheric pressure with absorption sensitivity of $6.2\times10^{-7}$ $cm^{-1}$ in 2 ms, demonstrating the potential of the technique for fast and sensitive measurement in the molecular fingerprint region. However, the simultaneous spectral coverage was limited by the arrangement of the fixed grating and the photodetector, and the entire spectral range was measured piecewise by manually rotating the grating, and the pieces were attached together in post processing. Moreover, the relative frequency jitter of the comb lines with respect to the cavity modes prevented averaging of the spectra, and frequency calibration was not performed, which precluded fitting of the theoretical model in order to quantify the gas concentrations.

Here we demonstrate an improved MIR CF-VS spectrometer based on the same DROPO that allows continuous acquisition of spectra in the entire signal range with efficient averaging and a calibrated frequency axis. The grating is now mounted on a galvo scanner that rotates as the cavity length is scanned, and the two scans are synchronized by an active lock to keep the selected Vernier order fixed on the detector. This tight active lock ensures good frequency reproducibility of consecutive spectra and thus allows long term averaging. The frequency scale is calibrated using a Fabry-Perot etalon or, alternatively, the galvo scanner position signal. The high quality of the experimental spectra enables multiline fitting of a model and retrieving gas concentrations from the fit with high accuracy and precision.

In the following, we first briefly recall the principles of the continuous-filtering Vernier spectroscopy to introduce the terms, equations and concepts needed in the rest of the paper. Afterwards we present the experimental setup and procedures, including the active lock and the frequency stabilization of the Vernier order, and the two methods used for frequency calibration of the spectra. In the results section we verify the performance of the spectrometer by fitting a broadband model to the spectrum of a calibrated $CH_4$ gas sample and retrieving the concentration from the fit. We evaluate the sensitivity and long term stability of the system using the Allan-Werle plot. Finally, we detect $CH_4$ and $H_2O$ in dry air and laboratory air, both at atmospheric pressure, and retrieve their concentrations by multiline fitting.

## 2. Principles of continuous-filtering Vernier spectroscopy

In continuous-filtering Vernier spectroscopy the cavity enhances the interaction length with the sample and filters the comb to enable sequential detection of the entire bandwidth using a photodiode. When the cavity free spectral range, $FSR_C = c/(2L)$, where $c$ is the speed of light and $L$ is the cavity length, is equal to an integer multiple of the laser repetition rate, $FSR_C = \mathrm{m} f_{\mathrm{rep}}$, every m$^{\mathrm{th}}$ comb line within the range allowed by the dispersion of the cavity mirrors is transmitted through the cavity, where m = 1, 2, 3… [31]. This condition is called the perfect match and corresponds to cavity length $L_{\mathrm{PM}} = c/(2\mathrm{m}f_{\mathrm{rep}})$. When the cavity length is detuned from $L_{\mathrm{PM}}$ by $\Delta L < L_{\mathrm{PM}}/F$, where $F$ is the cavity finesse, the cavity acts as a filter for the comb and transmits groups of comb lines called Vernier orders [26]. These orders are centered at frequencies $\nu_{\mathrm{k}} = c(\mathrm{k} - \delta f_0 / f_{\mathrm{rep}})/(2\mathrm{m}|\Delta L|)$ within the comb bandwidth, as shown in Fig. 1(a), where k is an integer number of the Vernier order and $\delta f_0$ is the mismatch between the comb and cavity offset frequencies [26]. The Vernier orders have a Lorentzian envelope defined by the peak transmission of the comb lines (red curve connecting the red markers). The separation of consecutive Vernier orders is given by $FSR_V = c/(2\mathrm{m}|\Delta L|)$, and their width (resolution) is given by $\Gamma_V = \mathrm{m}FSR_V / F = c/(2F|\Delta L|)$. In order to acquire a spectrum, the Vernier orders are spatially separated with a grating and the integrated intensity of a selected Vernier order is measured as it is tuned across the comb spectrum by scanning $\Delta L$. The tuning of the Vernier order can alternatively be obtained by scanning $f_{\mathrm{rep}}$, however, scanning $\Delta L$ is often preferred in practice as it can usually be done over a larger range than scanning the $f_{\mathrm{rep}}$. Nevertheless, the fact that the Vernier order frequency can be controlled via either $\Delta L$ or $f_{\mathrm{rep}}$ allows using both for stabilization of the Vernier order frequency, as described in section 3.1.

The resolution of the Vernier order changes as it is scanned across the spectrum because of the change of $\Delta L$. For instance, when the 70$^{\mathrm{th}}$ Vernier order is scanned across a spectrum spanning 0.25 μm around 3.25 μm, $\Delta L$ changes from 53.0 to 57.4 μm, which implies that the resolution changes by 8% during the scan (e.g. between 7.57-8.20 GHz for $F$ = 340). The variation of cavity finesse with wavelength also influences the resolution.

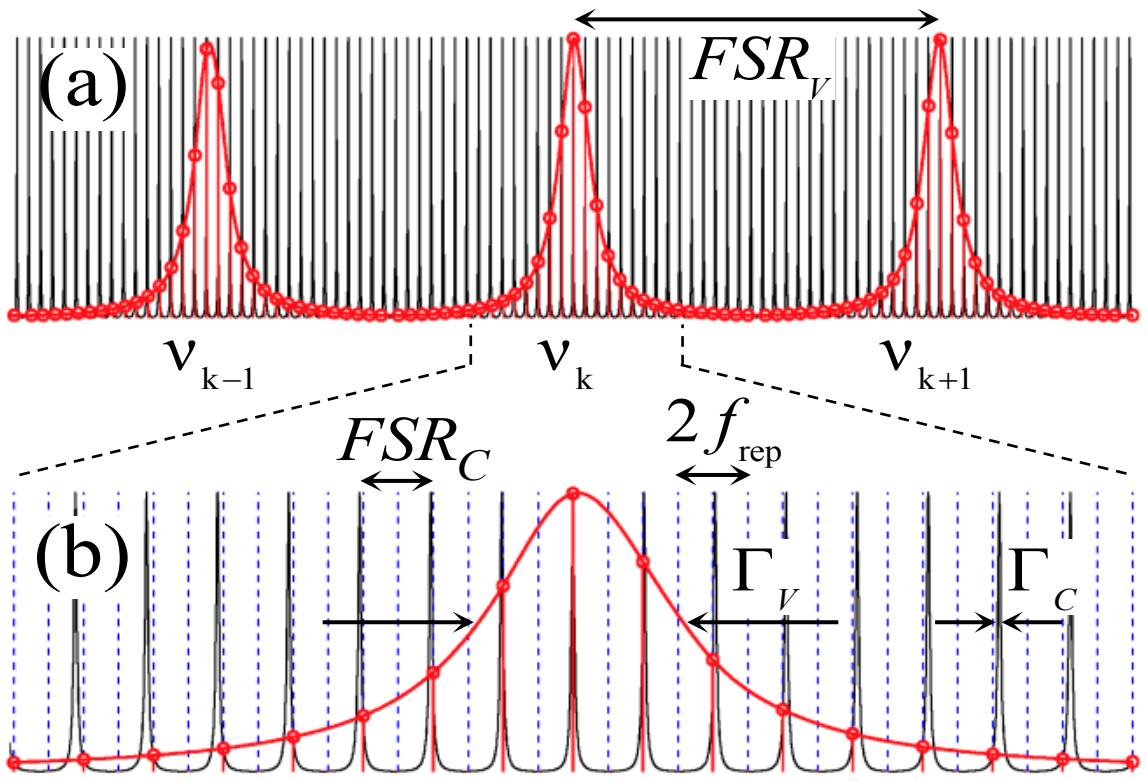

Fig. 1. (a) Three consecutive Vernier orders and (b) a zoom of the center order for $F = 20$, $\Delta L = -0.16L_{PM}$ and m = 2. The black curves are the cavity modes, the vertical blue dashed lines [shown in (b) but omitted in (a) for clarity] are the comb lines, while the vertical red lines are the comb lines transmitted through the cavity modes. The red markers indicate the intensities of the transmitted comb lines and the red curve is the envelope of the Vernier orders.

When the Vernier order containing very few comb lines is swept across the comb spectrum (in the absence of absorbing species) the measured intensity fluctuates. In order for this intensity modulation to stay below $10^{-6}$ the number of comb lines within the Vernier order, given by $N_V = \Gamma_V / (\mathrm{m}f_{\mathrm{rep}}) = L_{\mathrm{PM}} / (F|\Delta L|)$, should be at least 5 [26]. This implies that the lowest achievable resolution in the continuous-filtering limit is equal to $\Gamma_V^{\min} = 5FSR_C$ obtained when $\Delta L$ is equal to $|\Delta L^{\max}| = L_{\mathrm{PM}} / (5F)$. As an example, for a cavity with $L_{\mathrm{PM}} = 60$ cm and $F = 340$ this yields $\Gamma_V^{\min} = 1.25$ GHz for $|\Delta L^{\max}| = 350$ µm.

The maximum scanning speed of the Vernier order is limited by the cavity response time – to avoid distortion the cavity modes must be scanned adiabatically across the comb lines. This condition is given by $V_{\mathrm{sweep}}^{\max} = \Gamma_C^2 L_{\mathrm{PM}} / \Delta L = c^2 / (4L_{\mathrm{PM}}\Delta L F^2)$, where $\Gamma_C$ is the cavity mode width [26]. This implies that the 70$^{\mathrm{th}}$ Vernier order, for which $\Delta L = 55$ µm, can be scanned at $5.7\times10^{15}$ Hz/s in a cavity with $L_{\mathrm{PM}} = 60$ cm and $F = 340$. Thus scanning this order across a spectrum spanning 0.25 µm around 3.25 µm (which corresponds to $\Delta\nu = 7.1\times10^{12}$ Hz) takes at least 1.2 ms.

The Vernier transmission function for an empty cavity and in the presence of an absorbing gas has been described in detail in Ref. [26]. The intensity of one spectral element (i.e. Vernier order) is given by the sum of the intensities of all comb lines comprised in the range $\pm FSR_V/2$ around $\nu_k$ weighted by the Vernier order envelope (red curve in Fig. 1). In the presence of absorbing species, the cavity modes in the vicinity of a molecular transition are broadened, their amplitudes are reduced and their center frequencies are shifted away from the transition frequency. These three effects modify the profile of the Vernier order envelope (see Eq. 18 in [26]). The effect of amplitude reduction and mode broadening is the same for positive and negative values of $\Delta L$, while the shift of cavity resonance frequencies has a different effect depending on the sign of $\Delta L$. In the absence of absorption, when $\Delta L$ is negative the comb lines within a Vernier order are detuned from their respective cavity modes towards the center of the Vernier order [see Fig. 1(b)], while the opposite is true when $\Delta L$ is positive. In the presence of absorption, because of these different initial positions, the shift of the cavity modes caused by molecular dispersion induces a narrowing of the Vernier order in the negative case and a broadening of the order in the positive case. The final molecular Vernier spectrum, given by the ratio of Eq. 19 and 11 in [26], has a higher contrast for a negative than for a positive $\Delta L$ and the difference in contrast is larger when the Vernier order is broader than the absorption line. In the limit of low absorption, the molecular signal takes the same shape for both signs of $\Delta L$ and can be approximated by $1-FL\alpha/\pi$, where $\alpha$ is the molecular absorption [25].

## 3. Experimental setup and procedures

A schematic picture of the continuous-filtering Vernier spectrometer is shown in Fig. 2. The MIR comb source is a doubly resonant optical parametric oscillator (DROPO) based on an orientation-patterned GaAs (OP-GaAs) crystal (BAE Systems) pumped by a Tm:fiber femtosecond laser (IMRA America) with a repetition rate of 125 MHz, which delivers up to 2 W of power around 1.95 μm [10]. In the non-degenerate operation mode, the DROPO output (p-polarized) consists of a signal comb with ~250 nm bandwidth tunable between 3.1-3.6 μm and an idler comb with ~350 nm bandwidth tunable in the 4.6-5.4 μm range, providing up to 28 mW of signal and 20 mW of idler power [10].

The $f_{rep}$ of the pump laser is locked to the FSR of the DROPO cavity using the dither-and-lock method. The length of the DROPO cavity is dithered using a low amplitude sinewave signal at 12 kHz applied to a PZT on which one of the DROPO cavity mirrors is mounted, and the signal power is monitored using a PbSe detector. The output of the detector is demodulated by a lock-in amplifier (LiA, Stanford Research Systems, SR830) and the error signal is fed to a proportional-integral (PI) controller ('Slow $f_{rep}$ control', New Focus, LB1005) connected to a slow PZT acting on the $f_{rep}$ of the pump laser. The closed-loop bandwidth of the lock is 300 Hz, which is sufficient since the finesse of the DROPO cavity is quite low (~20). The carrier-envelope offset frequency ($f_{ceo}$) of the pump laser is free running.

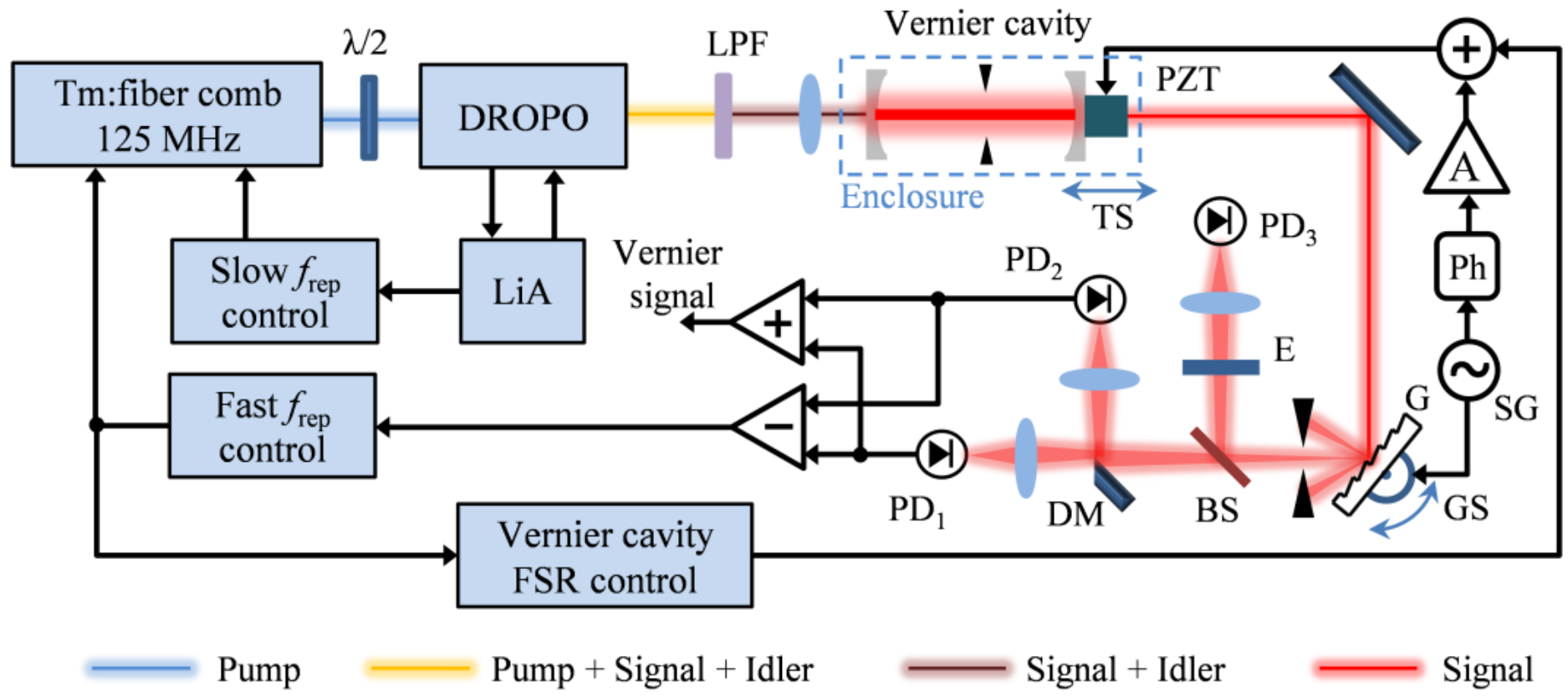

Fig. 2. Experimental setup: λ/2 – half-waveplate; DROPO – doubly resonant optical parametric oscillator; LiA – lock-in amplifier; LPF – long-pass filter; TS – translation stage; A – amplifier; Ph – phase shifter; SG – signal generator; GS – galvo scanner; G – diffraction grating; E – Fabry-Perot etalon; BS – beam splitter; DM – D-mirror; $PD_{1\text{-}3}$ – HgCdTe photodetectors.

The output of the DROPO is long-pass filtered to block the reminder of the pump power and the beam is mode-matched to the $TEM_{00}$ mode of the Vernier cavity using two lenses. The cavity consists of two dielectric concave mirrors with high reflectivity around 3.2 µm (LohnStar Optics, ZnSe substrate, 5 m radius of curvature) separated by 60 cm. The output mirror is mounted on a translation stage (TS) and attached to a PZT. The FSR of the Vernier cavity is equal to $2f_{rep}$, so at perfect match length, $L_{PM}$, every second signal comb line is transmitted through the cavity. The idler wavelength is not resonant in the cavity. An iris in the center of the cavity blocks the higher order transverse modes remaining after the mode-matching of the beam to the cavity, which are caused by the ellipticity of the signal beam originating from the angled concave mirrors in the DROPO cavity. Removing the transverse modes is crucial for the performance of the technique as they can overlap with the Vernier orders and thus distort the acquired spectrum by causing artificial absorption features. By closing the iris, the intensity of the strongest higher order transverse modes is reduced to ~1% of the peak intensity at the perfect matching point. The cavity is placed in an enclosure that can be filled with gas samples of calibrated $CH_4$, pure $N_2$ or dry air using flow controllers. For measurement of laboratory air we close the flow controllers and open the enclosure.

The wavelength-dependent cavity finesse ($F$) was determined by fitting Airy functions to cavity mode profiles measured in transmission. The cavity length was dithered around $L_{PM}$ by applying a 100 Hz sine wave to the cavity PZT and the $f_{ceo}$ of the signal comb was adjusted to match the cavity offset frequency by equalizing the intensities of the two neighboring transmission peaks around the perfect match condition. The transmitted light was dispersed by a reflection diffraction grating. By rotating the grating, different wavelength ranges of the transmission were imaged on a DC-coupled HgCdTe detector. Airy functions were fit to the transmission profiles measured at different wavelengths and the finesse was calculated from their linewidth. The results are shown in Fig. 3, where the black circular markers are the averaged values of 20 consecutive measurements and the error bars are the standard deviations. The average relative uncertainty in the cavity finesse measurement is 2%. The red curve is a fit of a 3$^{rd}$ order polynomial to the measured values, which is used later in the calculation of the Vernier spectrum model.

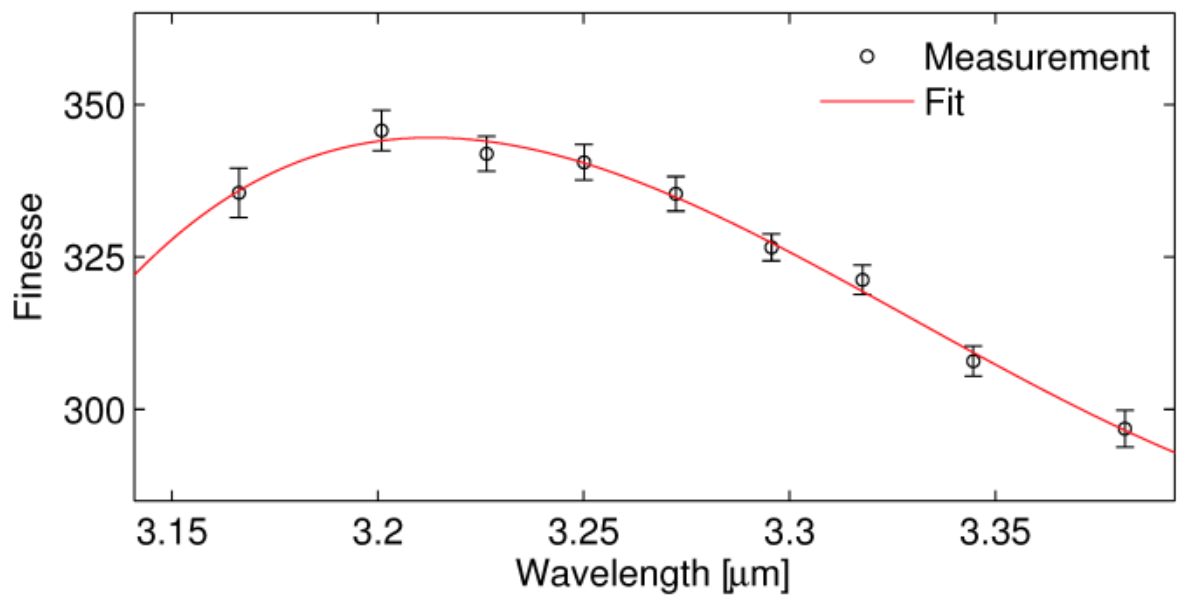

Fig. 3. Wavelength-dependent cavity finesse (black circular markers) along with a 3$^{rd}$ order polynomial fit (red curve). The standard deviation of each measurement is shown by the corresponding error bar.

For Vernier spectroscopy, the cavity length is detuned from $L_{PM}$ by $\Delta L$ using the translation stage. The optimum number of the Vernier order is selected as a compromise between the resolution and the signal-to-noise (SNR) ratio. A higher Vernier order yields better resolution but also lower transmitted power, as fewer comb lines are contained in the order. We operate the spectrometer at negative Vernier orders number 70-80, corresponding to $\Delta L$ between -55 and -63 µm, which contain 28-33 comb modes and yield 7.0-8.2 GHz resolution, sufficient for detection of $CH_4$ molecules at atmospheric pressure, which have linewidths of the order of 4 GHz. The optical power in the selected order after the grating is ~10 μW, which yields a maximum SNR on the Vernier signal of 650, limited by the detector noise.

The light transmitted through the cavity is incident on the reflection diffraction grating (Thorlabs, GR1325-45031, 450 grooves/mm) mounted on a galvo scanner (Thorlabs, GVS011) to spatially separate the different Vernier orders. The galvo scanner is dithered at 20 Hz by a sinewave from a low-noise signal generator (SG, Agilent, 33210A). The same signal is fed forward to the cavity PZT through a phase shifter (Ph) and an amplifier (A), to simultaneously scan the length of the cavity. An active lock, described in section 3.1, synchronizes the two scans in order to keep the chosen Vernier order fixed in space. An iris after the grating blocks the higher and lower Vernier orders from reaching the detection system. The beam is split using a pellicle beam splitter (Thorlabs, BP145B4) deployed around Brewster angle and a small portion (~20%) of the beam intensity is sent through a Fabry-Perot etalon (4.0 mm thick uncoated $CaF_2$ window, Thorlabs, WG50540) and focused on a DC-coupled HgCdTe detector ($PD_3$, Vigo Systems, PVI-4TE-6) for frequency calibration (described in section 3.2). The transmitted part of the beam is cut by a D-mirror and focused onto two DC-coupled HgCdTe detectors ($PD_1$ and $PD_2$, Vigo Systems, PVI-4TE-6). The difference of the outputs of the two detectors constitutes an error signal for the frequency locking, while the sum yields the Vernier signal. The spectrum is acquired during both directions of the galvo scan, each lasting 25 ms. The acquisition time is limited by the load on the galvo scanner that allows sweep frequencies up to 20 Hz, and it is longer than the limit set by the adiabatic scan condition, which is 1.2 ms for the same spectral coverage and Vernier order (see section 2). The Vernier spectrum and the Fabry-Perot etalon signal (alternatively the position signal of the galvo scanner, see section 3.2) are recorded with a 2-channel data acquisition card (National Instruments, PCI-5922) at 500 ksample/s and 24-bit resolution and saved for further processing.

### *3.1. Frequency stabilization of the Vernier order*

The simultaneous scanning of the grating and cavity length provides good initial spatial stability of the Vernier order, however an active lock is needed to compensate for nonlinearities in the galvo and cavity length scans and the dispersion of the cavity mirrors, which causes the FSR to change, as well as any drifts and fluctuations in the system. The error signal for the locking is obtained from the difference of the outputs of the detectors $PD_1$ and $PD_2$. When the beam is cut exactly in half by the D-mirror, both detectors see the same intensity and the difference of the signals is zero. When the beam moves horizontally, the signal on the two detectors becomes unbalanced, and the sign of the difference of the signals reflects the direction in which the beam moves. Thus the difference of the outputs of the two detectors constitutes an error signal for stabilization of the Vernier order in the vertical direction. Since the cavity PZT has a limited bandwidth and the pump laser is equipped with a much faster PZT, it is convenient to correct the large-amplitude low-frequency fluctuations in the system by feeding to the cavity length, while the low-amplitude high-frequency fluctuations by feeding to $f_{rep}$. Therefore the error signal is first fed via a PI controller ('Fast $f_{rep}$ control', New Focus, LB1005) to the fast PZT acting on the $f_{rep}$ of the pump laser, which provides 45 kHz of closed-loop bandwidth. The output of the first PI controller is sent to a second PI controller ('Vernier cavity FSR control', New Focus, LB1005) connected to the Vernier cavity PZT to correct for slower drifts in the system (250 Hz closed-loop bandwidth). The fast corrections applied to the $f_{rep}$ of the pump laser are not followed by the DROPO cavity length, however, they do not affect the intensity of the DROPO signal since the induced frequency shift, estimated to be ~500 kHz, is much smaller than the linewidth of the DROPO cavity (~6 MHz).

Figure 4(a) shows the open-loop error signal, i.e. the difference of the outputs of the two detectors, for three consecutive Vernier orders (around -70$^{th}$ order, $\Delta L \approx$ - 55 μm) recorded as the cavity length is scanned linearly and the grating position is fixed. The dips in the error signal are water absorption lines from the dry air sample in the cavity. The grating fully resolves the successive orders but does not resolve their width, $\Gamma_V$, therefore the zero-crossings of the open-loop error signals from the successive orders are separated by $FSR_V$ =

1.4 THz, while the separation of the extrema of one order is determined by the size of the beam hitting the D-mirror. Assuming a linear relationship between the time and the frequency domains, the slope of the error signal in the time domain, 920 V/s, can be recalculated to 1.4 μV/MHz in the frequency domain using the time separation between the zero-crossings of the error signals in the time domain, equal to 2.2 ms. Figure 4(b) shows the error signal when the feedback loop is closed. The standard deviation of the noise on the error signal is 200 μV, which translates into frequency stability of 140 MHz, calculated using the slope of the error signal. This implies that the frequency stability of the -70$^{th}$ order is equal to 2% of its resolution. This residual noise in the error signal is ~40% above the noise floor originating from the thermal noise of the detectors (140 μV) that sets the limit of the achievable frequency stability of the -70$^{th}$ Vernier order to 100 MHz in the current setup. Assuming a constant closed-loop noise level on the error signal, the absolute frequency stability is inversely proportional to the number of the Vernier order since the power per order, and thus the slope of the error signal, is inversely proportional to it. Under this condition, the relative frequency stability decreases quadratically with the order number. Improving the frequency stability for a particular order would require increasing the optical power per comb mode or increasing the grating resolution in order to increase the slope of the error signal.

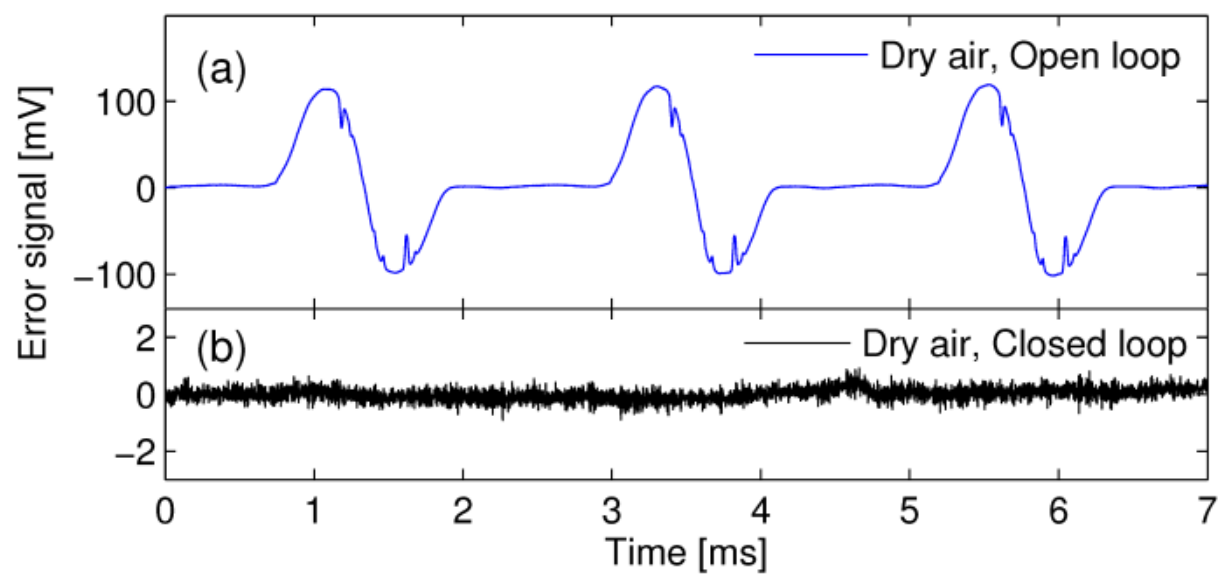

Fig. 4. (a) Open-loop error signal for three consecutive Vernier orders (around the -70$^{th}$ order) recorded as the cavity length is scanned linearly while the grating position is fixed. (b) Closed-loop error signal while the spectrometer is locked to the -70$^{th}$ order and the grating position is fixed.

Figure 5(a) and (c) show the open-loop error signal when both the grating and the Vernier cavity length are scanned and the cavity is filled with dry air (a) or laboratory air (c). The feedforward parameters are adjusted so that the two scans are almost synchronized and the selected Vernier order is kept on the detectors for a time much longer compared to when the grating is not scanned (compare the time scales in Fig. 4 and Fig. 5). The fact that a larger spectral range is imaged on the detectors during the scan is also reflected by the appearance of more absorption lines. Figure 5(b) shows the closed-loop error signal when the cavity is filled with dry air and the grating is scanned. The standard deviation of the noise is 250 μV, translating to a frequency stability of 170 MHz, which is slightly worse compared to when the grating position is fixed. When the cavity is filled with laboratory air [Fig. 5(c) and (d)], the large water absorption decreases the SNR of the error signal at the positions of strong absorption lines. This causes local loss of the locking and worse frequency stability at the positions of these lines, as can be seen in the closed-loop error signal in Fig. 5(d). The standard deviation of noise reaches 1.0 mV at the position of the strong absorption lines, corresponding to a frequency stability of 700 MHz. This effect causes a systematic distortion of the lineshape of the strong absorption lines in the Vernier spectrum.

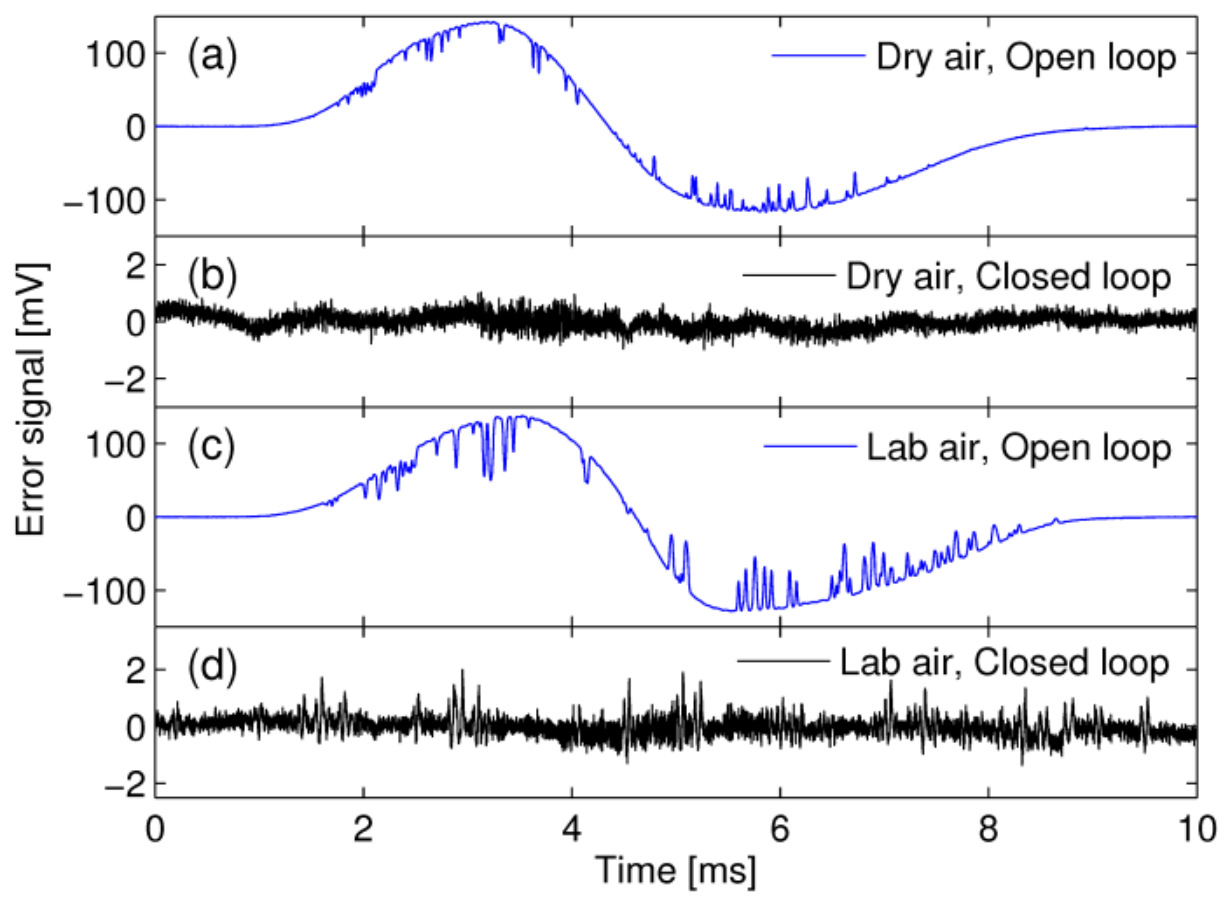

Fig. 5. (a) and (c) Open-loop error signals for -70$^{th}$ Vernier order recorded as the cavity length and the grating position are scanned simultaneously but slightly out of synchronization, when the cavity is filled with dry air (a) and laboratory air (c), respectively. (b) and (d) The corresponding closed-loop error signals for the -70$^{th}$ Vernier order recorded as the cavity length is scanned.

### *3.2. Frequency calibration*

The relative frequency scale of the Vernier spectrum is calibrated using the Fabry-Perot etalon signal from $PD_3$. To remove absorption lines from the etalon signal, the output of detector $PD_3$ is rescaled to correct for the different detector gains and divided by the Vernier signal. The normalized etalon signal is low-pass filtered to remove high frequency noise and the zero crossings of the fringes provide frequency markers separated by the FSR of the etalon, which is around 27 GHz. To keep the number of data points in the spectrum constant the frequency scale and the spectrum are linearly interpolated between these markers. The etalon FSR is determined by the thickness of the etalon and the wavelength dependent refractive index of $CaF_2$ calculated using the Sellmeier equation [32]. The absolute frequency scale is found by comparing the line positions in the spectrum to that listed in the HITRAN database [33]. In order to match these, we correct the wavelength dependence of the etalon FSR using a small 2$^{nd}$ order polynomial function. This function was found to be constant from day to day, which confirms that it corrects for an inaccuracy in the dispersion relation rather than for some random error. This implies that the relative frequency calibration is reproducible and the absolute scale can be found each time by a simple shift to match the positions of the absorption lines to that listed in HITRAN.

In case of strong molecular absorption (cavity transmission <70% at the peak of the absorption line) the etalon fringes are strongly attenuated and their zero crossings cannot be retrieved correctly. For such samples, the frequency scale is instead calibrated using the position signal of the galvo scanner. The position signal is low-pass filtered to minimize the noise and pre-calibrated using the etalon signal from an $N_2$-filled cavity. The absolute frequency scale is again found by comparison to line positions listen in HITRAN. However, this time a 4$^{th}$ order polynomial correction function is needed to compensate the remaining discrepancies, which originate mainly from the alignment of the beam on the grating, and it was found to vary significantly from day to day. Thus calibration by the etalon is preferred whenever possible.

## 4. Results

### *4.1. Frequency and amplitude stability*

To demonstrate the stability of the spectrometer, we fill the cavity enclosure with dry air and measure the Vernier signal. Figure 6 shows ten consecutive Vernier spectra measured using the -70$^{th}$ order with frequency scale calibrated using the etalon signal, demonstrating that the overall frequency and intensity reproducibility of the spectra is very good. The inset shows a zoom of the center of one of the $CH_4$ absorption lines. We find the center frequency of this absorption line in each spectrum by fitting a 9$^{th}$ order polynomial to the data shown in the inset, and the standard deviation of these frequencies is 180 MHz, which is consistent with the frequency stability of 170 MHz determined from the noise in the error signal shown in Fig. 5(b). This implies that the relative frequency stability is 2% of the linewidth of the absorption lines in the Vernier spectrum (10 GHz), which allows efficient averaging of the spectra with negligible distortion on the absorption features.

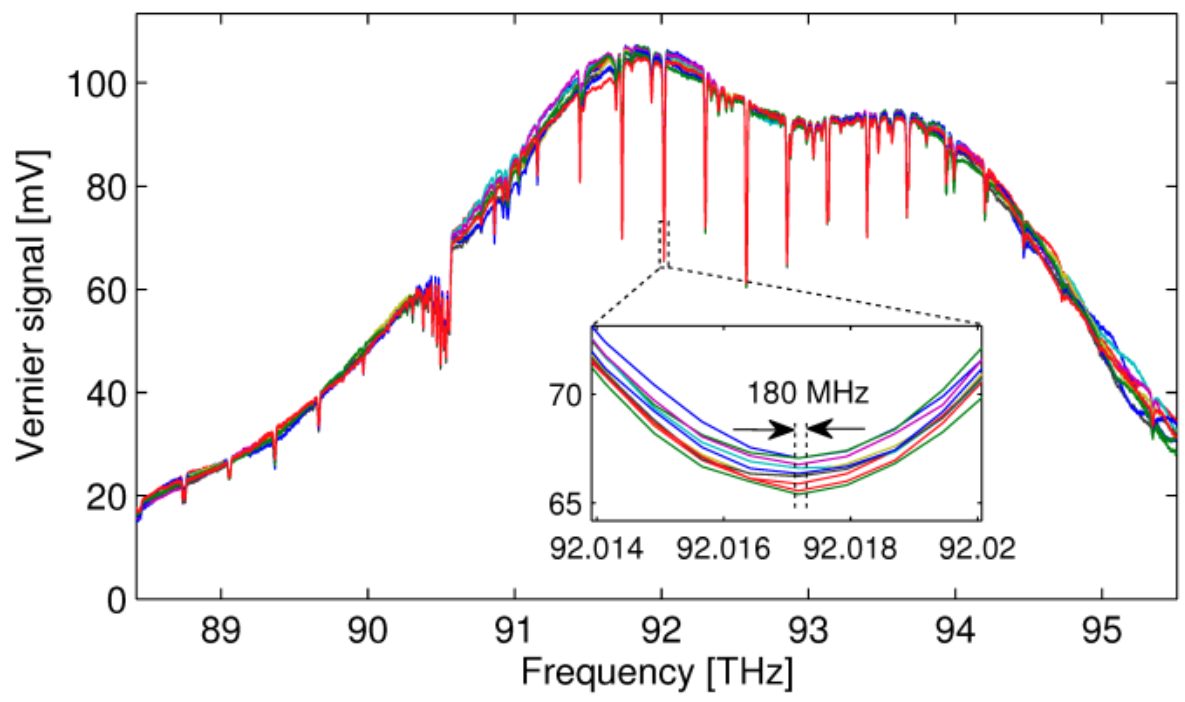

Fig. 6. Ten consecutive Vernier spectra of dry air with frequency scale calibrated using the etalon. The inset shows a zoom of the peak of a $CH_4$ absorption line with standard deviation of the center frequency from consecutive measurements.

### *4.2. Spectral fitting and concentration retrieval*

Figure 7(a) shows in black the Vernier spectrum of a calibrated sample of 5.00(5) ppm of $CH_4$ in $N_2$ at atmospheric pressure normalized to a background spectrum measured when the cavity is filled with pure $N_2$ and averaged 10 times. The red curve shows a fit of a $CH_4$ Vernier spectrum with concentration as the fitting parameter, calculated for the -69$^{th}$ order with line parameters from the HITRAN database and the wavelength dependent finesse of the cavity shown in Fig. 3. A sum of a third order polynomial and low frequency etalon fringes is fitted and subtracted from the data to compensate for the drift of the baseline. The residual of the fit is shown in Fig. 7(b), demonstrating that the general agreement between the measured spectrum and the model is good. Two different absorption features [indicated by '*' and '+' in (a)] are enlarged in Fig. 7(c) and (e) (black markers) along with the corresponding fit (red curves) and the residuals of the fits are shown in Fig. 7(d) and (f). The structure remaining in the residuals is mainly caused by the remaining error in the frequency calibration and possibly by the different gains and nonlinearities in the response of the two detectors used for recording the spectrum ($PD_1$ and $PD_2$). The concentration retrieved from the fit is 5.00(3) ppm, where the error is the standard deviation of 10 consecutive measurements. This result agrees with the specified sample concentration, which shows that the accuracy of the spectrometer is better than 1% in 250 ms.

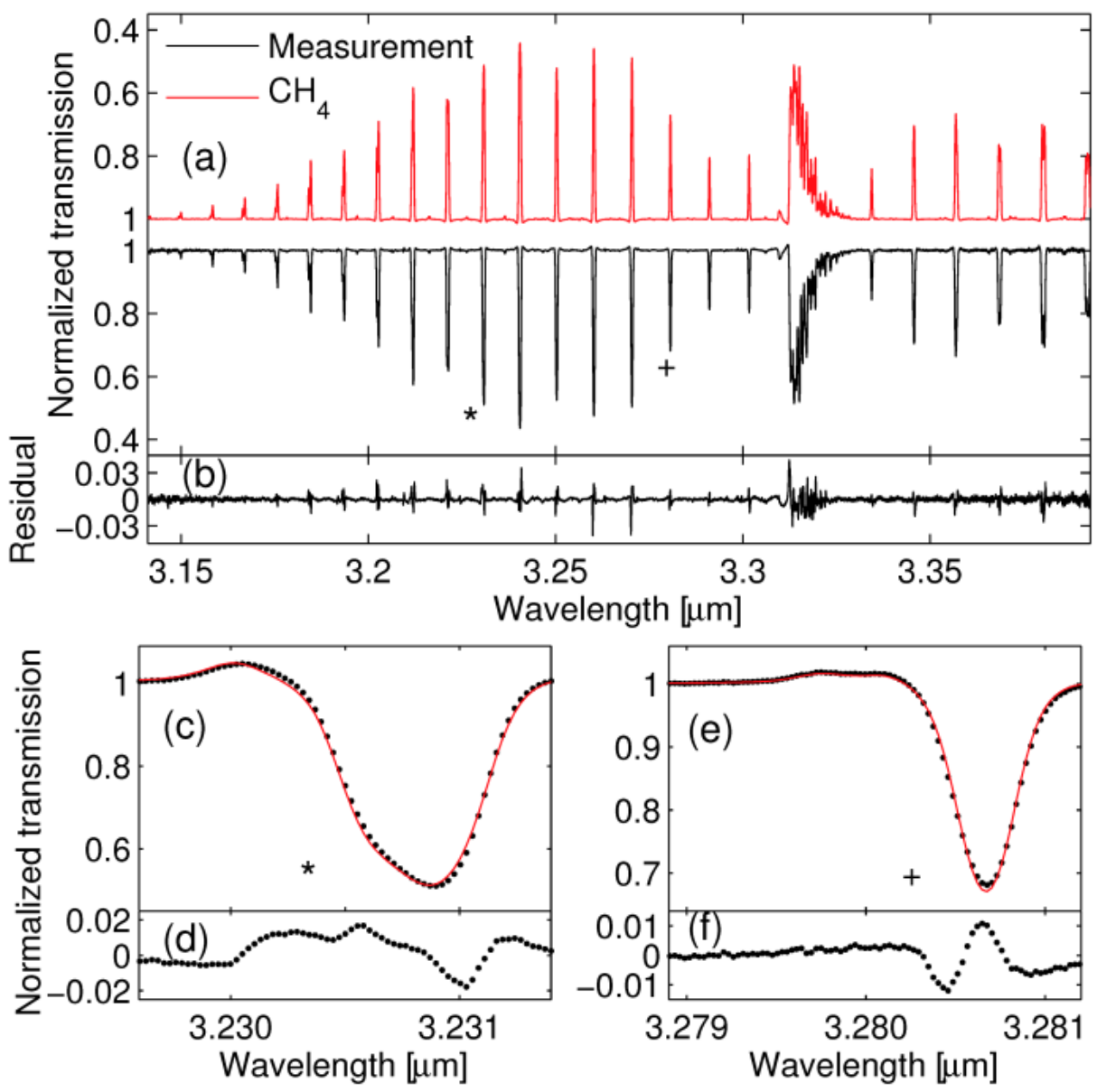

Fig. 7. (a) Normalized Vernier spectrum of 5 ppm of $CH_4$ in $N_2$ at atmospheric pressure (black, 8.3 GHz resolution, 10 averages, 250 ms) along with a fitted model (red, inverted for clarity). (b) Residual of the fit. (c) and (e) Zooms of two different absorption features [black markers, indicated by '*' and '+' in (a)] along with the corresponding fits (red curves). (d) and (f) The residuals of the fits.

It should be noted that determining the exact number of the Vernier order from $\Delta L$ is practically impossible as $\Delta L$ changes by 0.8 μm between two consecutive orders. Therefore we detune the cavity length by approximately the desired value of $\Delta L$ from $L_{PM}$ and determine the exact number of the Vernier order by fitting a model to the spectrum with both concentration and the Vernier order number as fitting parameters. Next, we round the Vernier order number found from the fit to the closest integer number and we fix it in the fits to spectra recorded while the system stayed locked.

The high quality of the spectra and the accuracy of the theoretical model allow extracting gas concentration from samples containing many species using multiline fitting. This is demonstrated in Fig. 8(a), which shows a Vernier spectrum of dry air at atmospheric pressure (black) together with the fitted model spectra of $CH_4$ and $H_2O$ spectra (red and blue, respectively). The concentrations found from the fit are 2.007(9) ppm of $CH_4$ and 113.3(5) ppm of $H_2O$, and the $CH_4$ concentration agrees with that expected in air. The residual of the fit is shown in Fig. 8(b). For comparison, the same two spectral ranges as in Fig. 7, now containing both $CH_4$ and $H_2O$ lines, are enlarged in Fig. 8(c) and (e) (black markers) along with the corresponding fits ($CH_4$ in red and $H_2O$ in blue) and the residuals of the fits are shown in Fig. 8(d) and (f). The structures in the residuum have lower amplitudes because the absorption in the spectrum is lower, which improves both the frequency stability and calibration. Even though the resolution is ~8.2 GHz, which is larger than the molecular linewidths, it is possible to determine the concentration of the two different species with good precision in a short measurement time by simultaneously fitting the models of both spectra.

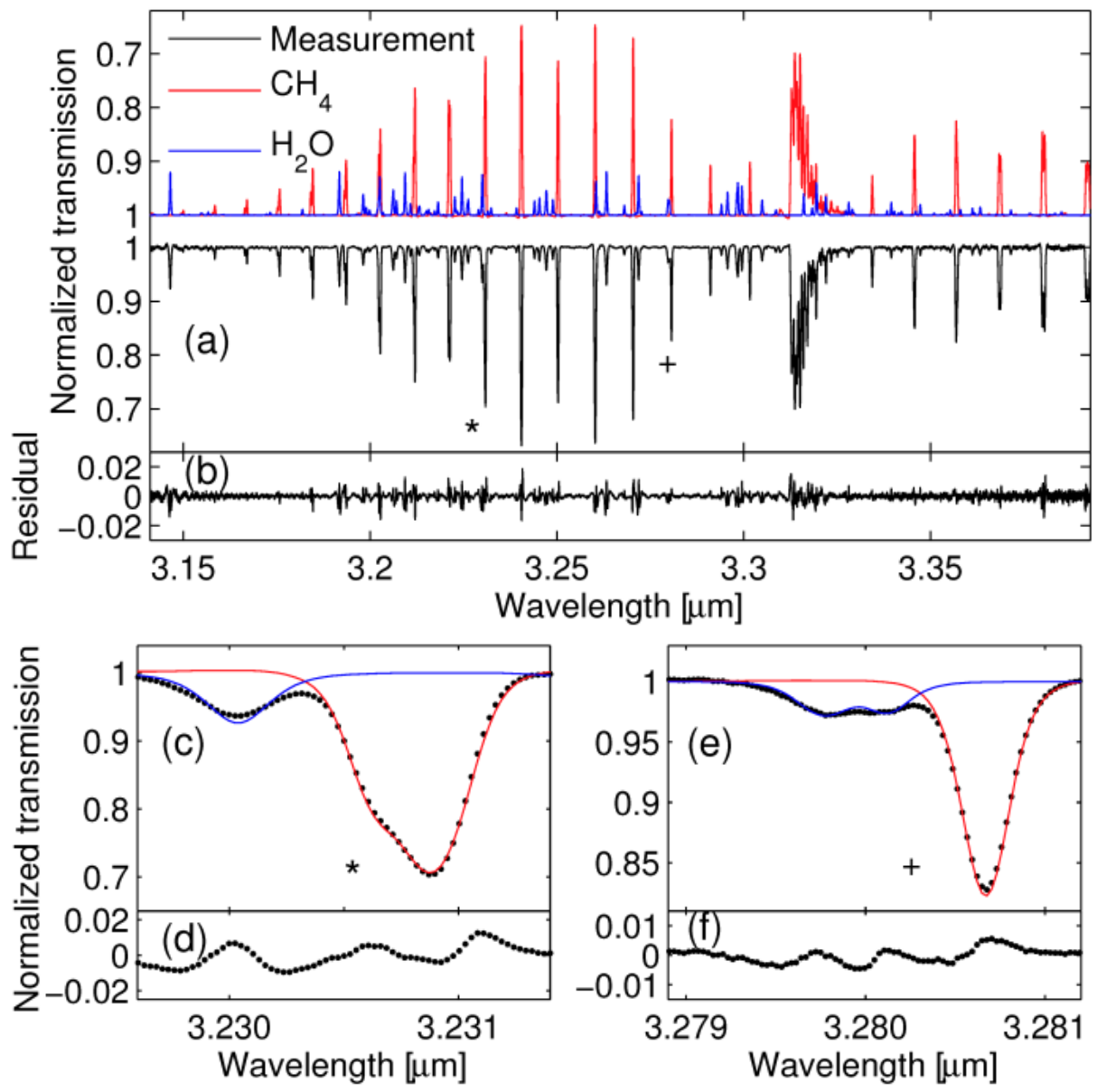

Fig. 8. (a) Normalized Vernier spectrum of dry air at atmospheric pressure (black, 8.2 GHz resolution, 10 averages, 250 ms) along with a fitted spectrum of $CH_4$ (red) and $H_2O$ (blue). (b) Residual of the fit. (c) and (e) Zooms of two different absorption features [black markers, indicated by '*' and '+' in (a)] along with the corresponding fits ($CH_4$ in red and $H_2O$ in blue). (d) and (f) The residuals of the fits.

### *4.3. Sensitivity and long term stability*

We estimate the absorption sensitivity by taking the ratio of two consecutive background spectra (measured when the cavity is filled with $N_2$) and fit and remove the baseline as in the treatment of the spectra described above. The standard deviation of the noise at ~3.2 μm is found to be $\sigma = 1.5\times10^{-3}$ which corresponds to a noise equivalent absorption coefficient of $5.2\times10^{-8}$ cm$^{-1}$ Hz$^{-1/2}$, calculated as $\sigma T^{1/2}/L_{eff}$, where $T$ is the acquisition time of two spectra and $L_{eff}$ is the effective length, defined as $FL/\pi$, with $F$ = 340 at 3.2 μm. We define the number of resolved elements, $M$, as the ratio between the entire spectral range (7.2 THz or 240 $cm^{-1}$) and the resolution of the spectrometer (8.2 GHz or 0.273 $cm^{-1}$), which yields 880, thus the figure of merit, calculated as $\sigma/L_{\mathrm{eff}}(T/M)^{1/2}$, is equal to $1.7\times10^{-9}$ $cm^{-1}$ $Hz^{-1/2}$.

To estimate the concentration detection limit and long term stability of the spectrometer we measure background spectra with the cavity filled with $N_2$ for 20 minutes and normalize each of them to the first spectrum. Afterwards we fit a sum of the model of $CH_4$ Vernier spectrum and a slowly varying baseline to these normalized spectra with concentration as the fitting parameter. The Allan–Werle plot of the concentrations found from these fits is shown in Fig. 9 (black). The dashed red line shows the $\tau^{-1/2}$ dependence characteristic for the white-noise-dominated regime that is fitted to the corresponding measurement points. The $CH_4$ concentration detection limit, given by the slope of the fitted line, is 360 ppt $Hz^{-1/2}$. The Allan–Werle plot shows that the absolute minimum detectable concentration of $CH_4$ will be ~90 ppt after 16 s of integration.

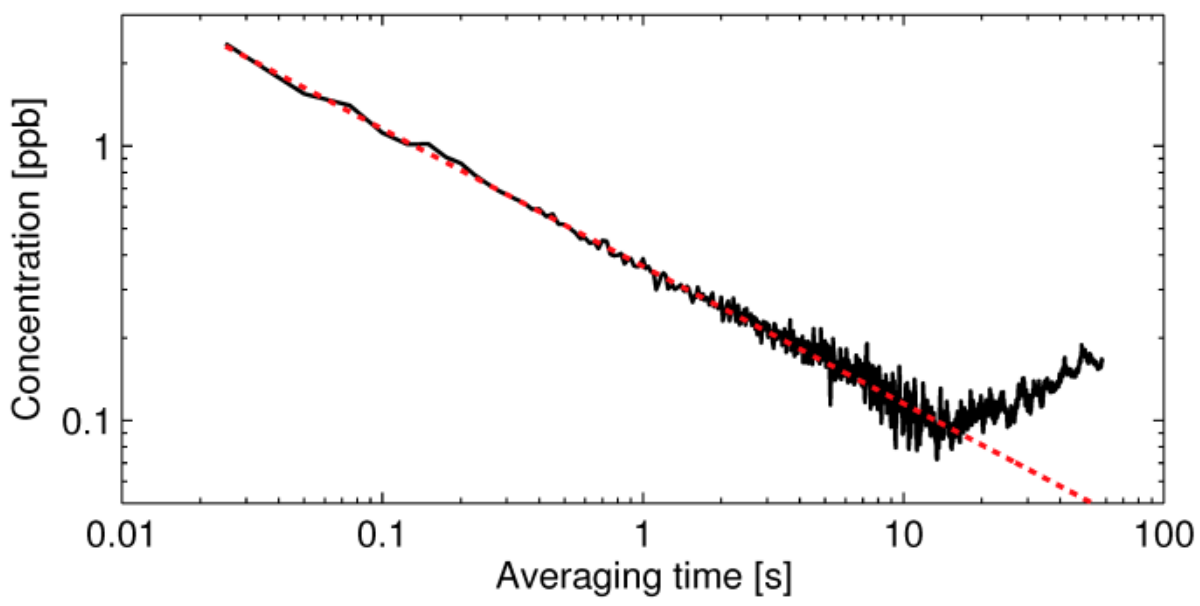

Fig. 9. Allan–Werle plot of the minimum detectable $CH_4$ concentration retrieved from fitting of the $CH_4$ Vernier spectra to normalized background spectra (black) and the linear fit to the white-noise-dominated regime (dashed red).

### *4.4. Spectra of strongly absorbing species*

As mentioned in sections 3.1 and 3.2, strong absorption lines not only degrade the locking performance and frequency stability, but also attenuate the etalon signal and prevent correct retrieval of its zero crossings for frequency calibration. This precludes the use of the etalon signal for frequency calibration. In order to enable measurements of strongly absorbing samples, we use the position signal of the galvo scanner for the relative frequency calibration. A normalized Vernier spectrum of laboratory air at atmospheric pressure, measured using the -84$^{th}$ order and calibrated using the galvo scanner position signal, is shown in black in Fig. 10(a). The red and blue inverted curves show the fitted $CH_4$ and $H_2O$ spectra, respectively, with the concentrations as the fitting parameter. The residual of the fit is shown in Fig. 10(b). The concentrations found from the fit are 2.08(1) ppm of $CH_4$ and 0.641(1)% of $H_2O$. Two different absorption features [indicated by '*' and '+' in (a)] are enlarged in Fig. 10(c) and (e) (black markers) along with the corresponding fits ($CH_4$ in red and $H_2O$ in blue) and the residual of the fits are shown in Fig. 10 (d) and (f). The general agreement between the measured spectrum and the model is still good, but the residual of the fit has larger amplitude and is more structured than in the spectra of dry air calibrated using the etalon. The main reason is the distortion and shift of the strong absorption lines caused by the local loss of the frequency lock, which can be easily identified in Fig. 10(c). Nevertheless, even though the spectra have worse frequency precision compared to the etalon calibrated spectra, it is still possible to discriminate the two species by multiline fitting and determine their concentrations with good precision. The local distortions of the strongest lines are in fact averaged out by the multiline fitting procedure, and the larger area of water absorption yields good precision in concentration determination [3].

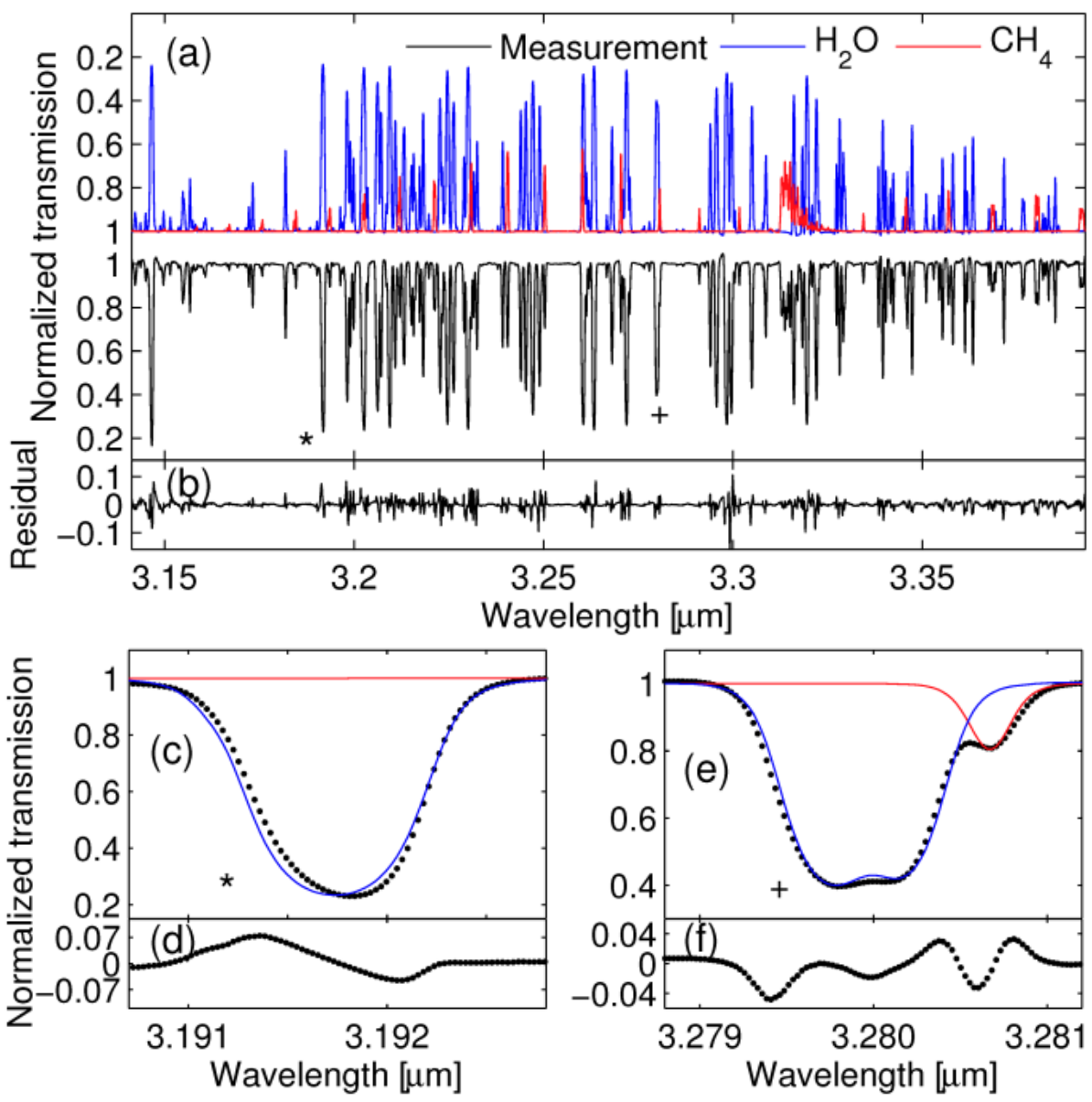

Fig. 10. (a) Normalized Vernier spectrum of laboratory air at atmospheric pressure (black, 6.8 GHz resolution, 10 averages, 250 ms) with frequency scale calibrated using the galvo scanner position signal, along with a fitted model spectrum of $CH_4$ (in red) and $H_2O$ (in blue). (b) Residual of the fit. (c) and (e) Zooms of two different absorption features [black markers, indicated by '*' and '+' in (a)] with the corresponding fits ($CH_4$ in red and $H_2O$ in blue). (d) and (f) The residuals of the fits.

## 5. Conclusions and outlook

We demonstrated a MIR continuous-filtering Vernier spectrometer based on a doubly resonant optical parametric oscillator capable of acquiring a spectrum in the entire span of the signal comb (250 nm around 3.25 μm) with a resolution of 8 GHz in 25 ms. The active lock that synchronizes the scan of the grating and the cavity, and thus keeps the selected Vernier order fixed on the detectors, enables uninterrupted continuous operation and efficient averaging of the consecutive spectra over almost 3 decades. We implemented and compared two different methods of relative frequency scale calibration, using a Fabry-Perot etalon or the position signal of the galvo scanner. The former method offers better precision and long term reproducibility, but cannot be used for highly absorbing samples, as the etalon signal is lost at the positions of strong absorption lines. Thus calibration using the galvo scanner position is performed for the highly absorbing samples. The frequency stability was estimated to be 170 MHz, a fraction of the resolution of the spectrometer, limited by the detector noise in the error signal. We used the model of the Vernier signal to perform multiline fitting to $CH_4$ and $H_2O$ spectra and we retrieved the gas concentrations with high accuracy and precision (~1%) in 250 ms. The figure of merit of the spectrometer is $1.7\times10^{-9}$ cm$^{-1}$ Hz$^{-1/2}$ per spectral element, and the minimum detection limit for $CH_4$, estimated by multiline fitting and the Allan-Werle plot, is 360 ppt Hz$^{-1/2}$, averaging down to ~90 ppt at 16 s.

The sensitivity of our CF-VS spectrometer is detector noise limited, so it can be improved by increasing the power per comb line, increasing the cavity finesse, and using a cavity with an $FSR_C$ matched to $f_{\mathrm{rep}}$ rather than $2f_{\mathrm{rep}}$. The frequency stability of the system would also improve if more power was available as it would increase the SNR in the error signal used for locking. Higher power per comb line would also allow working on a higher Vernier order to increase the resolution and allow measurements at lower pressures. The acquisition time is currently limited by the load of the galvo scanner and is a factor of 20 longer than the fundamental limit, which means that working on a higher Vernier order is possible without increasing the scan time. Thus it can be concluded that the performance of our spectrometer is mainly limited by the available optical power provided by our MIR comb source. Using the DROPO requires an additional servo loop to stabilize the pump and OPO cavities to each other, which adds some complexity to the system. Moreover, the available reflectivity range of dielectric cavity mirrors does not currently allow simultaneous measurement over the entire bandwidth of the DROPO, which implies that another pair of mirrors must be used to perform measurements in the idler comb range (4.6-5.4 μm). Such cavity will be implemented in the future to allow simultaneous detection of molecular species such as NO, $N_2O$, $O_3$ and CO. For measurements around 3.3 μm we plan to implement the CF-VS spectrometer with a difference frequency generation (DFG) source recently developed in our lab that provides 200 mW of power in 200 nm of bandwidth around this wavelength. The DFG comb does not incorporate a cavity, which will simplify the locking.

In conclusion, continuous-filtering Vernier spectroscopy is a broadband, highly sensitive, and fast spectroscopic technique that allows precise quantification of concentrations of different species in the sample. The comb intensity is measured in cavity transmission using a compact and robust setup consisting of a grating, a D-mirror and two photodiodes. The provided spectral coverage, limited only by the reflectivity of the cavity mirrors, is the widest offered by any cavity-enhanced comb-based technique. The use of high finesse cavity together with the inherent noise immunity of the technique yields high absorption sensitivity and thus low concentration detection limits. Thus the continuous-filtering Vernier spectroscopy offers a viable alternative to other detection methods of MIR comb spectroscopy, particularly suited for trace gas detection.

**Acknowledgements**

This project was supported by the Swedish Research Council (621-2012-3650), Swedish Foundation for Strategic Research (ICA12-0031), Knut and Alice Wallenberg Foundation (KAW 2015.0159), and the Faculty of Science and Technology, Umeå University.